**Annealing-atmosphere control of paramagnetic defects in ZnO microparticles**

Dariya Savchenko[1,2,*], Maryna Holiatkina[1], Andrey Prokhorov[4], Andrii Vasin[3], Ekaterina Kalabukhova[3]

[1]*National Technical University of Ukraine "Igor Sikorsky Kyiv Polytechnic Institute", Kyiv, 03056, Ukraine*

[2]*Technical Center NAS of Ukraine, Kyiv, 04070, Ukraine*

[3]*V.Ye. Lashkaryov Institute of Semiconductor Physics of the National Academy of Sciences of Ukraine, Kyiv 03028, Ukraine*

[4]*Institute of Physics of the CAS, 18200, Prague 8, Czech Republic*

**ABSTRACT**

In this work, we used temperature-dependent electron paramagnetic resonance (EPR) spectroscopy to reveal the effect of high-temperature annealing in $N_2$ and air on paramagnetic and electronic properties of ZnO microparticles. EPR spectral simulations have shown that $N_2$ annealing produces $Zn_i$, $V_O^+$, and shallow-donor centers, whereas $Zn_i$, $V_O^+$, and $H_i$-related centers are observed after annealing the ZnO microparticles in an air atmosphere. The resolved paramagnetic centers exhibit distinctly different behavior over 5-297 K. The shallow-donor and $H_i$ signals are rapidly quenched below approximately 50 K, while the $Zn_i$ contribution shows a nonmonotonic temperature dependence. In contrast, the $V_O^+$ signal persists up to room temperature and increases with temperature, consistent with charge redistribution among donor-related defects involving deep oxygen-vacancy states. The cavity *Q*-factor generally decreases with increasing temperature for both annealed samples, indicating atmosphere-dependent changes in the microwave losses. These results show that the annealing atmosphere controls the paramagnetic defect population, the temperature dependence of donor-related centers, and the associated electronic response of ZnO microparticles.

*Corresponding author: Assoc. Prof., Dr.Sc. Dariya Savchenko
E-mail: d.v.savchenko@kpi.ua
Full postal address: Department of General Physics and Modelling of Physical Processes, Faculty of Physics and Mathematics, National Technical University of Ukraine "Igor Sikorsky Kyiv Polytechnic Institute", pr. Beresteiskyi 37, Kyiv, 03056, Ukraine

## 1. Introduction

Zinc oxide (ZnO) is a wide-band-gap II-VI semiconductor that combines promising electrical, optical, piezoelectric, and chemical properties, making it attractive for a broad range of electronic and optoelectronic applications [1,2]. ZnO has a direct band gap of about 3.37 eV and a relatively high exciton binding energy of about 60 meV, enabling efficient excitonic processes even at room temperature [1]. In addition, the high optical transparency in the visible range, chemical and thermal stability, and ability to control electrical conductivity make ZnO suitable for ultraviolet photodetectors and light-emitting devices, transparent conducting layers, sensors, and other functional devices [1-3].

Researchers have paid particular attention to micro- and nanostructured ZnO because particle size, morphology, surface states, and structural defects can strongly affect their functional properties [2-4]. ZnO micro- and nanostructures have been investigated for gas sensing, photocatalytic, optoelectronic, environmental, and biomedical applications, where their performance depends on structural and defect properties [3,4]. ZnO particles can also serve as precursors for fabricating functional films and coatings; therefore, controlling their defect and electronic properties before further processing is important [1]. Native and impurity-related defects can substantially modify the electronic structure of ZnO and consequently its electrical conductivity, optical response, and surface activity [4]. In ZnO microparticles, donor-related paramagnetic defects have also been shown to correlate with enhanced n-type conductivity, underscoring the need to identify their microscopic origin and thermal behavior [5].

The physical properties of ZnO are strongly influenced by native point defects and unintentional impurities, which can introduce donor- or acceptor-related electronic states and modify charge-carrier transport, optical response, and other functional properties [6,7]. The commonly considered intrinsic defects include oxygen vacancies ($V_O$), zinc vacancies ($V_{Zn}$), zinc interstitials ($Zn_i$), and oxygen interstitials ($O_i$). Despite extensive experimental and theoretical investigations, the microscopic origin of the n-type conductivity commonly observed in nominally undoped ZnO remains under discussion. In particular, $V_O$ are often associated with oxygen deficiency and donor-like behavior. However, $V_O$ are generally considered deep donors and are therefore unlikely to be the primary source of the high free-electron concentration in n-type ZnO [6]. Hydrogen-related defects and $Zn_i$ have also been considered as possible donor centers in ZnO [8,9]. Experimental EPR studies of ZnO microparticles have resolved several donor-related paramagnetic centers attributed to $V_O^+$, $Zn_i$, hydrogen interstitials ($H_i$), and shallow donors, showing that different defect species can coexist and contribute differently to the electronic properties of the material [5].

Thermal annealing effectively modifies ZnO defect structure because the formation, charge state, and stability of point defects depend strongly on temperature and chemical environment [10,11]. In particular, changing the annealing atmosphere alters the oxygen chemical potential and can therefore promote or suppress different intrinsic and impurity-related defect configurations. Oxygen-rich and oxygen-poor conditions can favor different defect configurations, while reactive annealing atmospheres may additionally modify impurity-related centers [10,11]. Recent studies show that thermal treatment under different atmospheres produces pronounced changes in the structural and paramagnetic defect populations of ZnO. For example, studies of ZnO nanoparticles annealed in air, oxygen, and argon revealed atmosphere-dependent changes in paramagnetic defect concentrations and characteristics [11]. These observations show that the annealing atmosphere is a key parameter for controlling the ZnO defect structure and point to the need for defect-sensitive spectroscopic methods that can distinguish different defect species.

Electron paramagnetic resonance (EPR) spectroscopy is a powerful tool for microscopic characterization of electrically active defects in ZnO because it provides direct information on paramagnetic centers with unpaired electrons. Analysis of the *g*-factor, spectral lineshape, and temperature dependence of the EPR signal can distinguish defect species and their charge states and provide insight into electron localization and thermal activation processes [5,7,13]. Previous EPR studies have revealed various intrinsic and impurity-related paramagnetic centers in ZnO, including signals associated with oxygen vacancies, zinc interstitials, hydrogen-related defects, and shallow donors [8,9]. Moreover, temperature-dependent EPR measurements can reveal the thermal stability and ionization behavior of donor-related centers in ZnO, allowing distinction between shallow and deeper donor-related states.

Despite considerable progress in understanding ambient-dependent defect formation in ZnO, the properties of individual paramagnetic centers produced under different annealing conditions remain insufficiently understood. The EPR studies in [11] show that the annealing atmosphere modifies the paramagnetic defect population of ZnO; however, little attention has been given to the thermal evolution of individual defect-related EPR components over a broad temperature range. Such information is important for distinguishing weakly localized donor states from more thermally stable defect centers and for understanding charge redistribution between them. In this work, we use temperature-dependent EPR spectroscopy to investigate ZnO microparticles annealed at high temperatures in air and nitrogen ($N_2$) atmospheres. By combining spectral simulations with analysis of the corrected double-integrated intensities over the 5-297 K range, we resolve the temperature evolution of individual donor-related paramagnetic centers. We pay particular attention to $Zn_i$, $H_i$, shallow-donor, and $V_O^+$-related centers and the effect of the annealing atmosphere on their populations and thermal behavior. In addition, the temperature dependence of the cavity *Q*-factor was analyzed to assess atmosphere-dependent changes in the microwave electrical response of ZnO. The results show that the annealing atmosphere governs not only the population of paramagnetic defects but also their thermal behavior, providing microscopic insight into ambient-induced changes in the defect and electronic structure of ZnO microparticles.

## 2. Materials and methods

Commercial ZnO microparticles with a nominal average grain size of approximately 5 μm were used as the starting material. Three sample types were investigated: as-received ZnO microparticles and ZnO microparticles annealed in air and $N_2$. For $N_2$ annealing, ZnO powder in a quartz boat was placed in a quartz tube of the furnace. To reduce air contamination, we evacuated the furnace to approximately 10 Pa using a mechanical vacuum pump, then filled it with pure $N_2$ to atmospheric pressure. Then, the $N_2$ flow rate was set to 100 sccm and kept constant during the heating-annealing-cooling process. The thermal treatment was performed at 1000 °C for 30 min. The as-received ZnO microparticles were used as a reference sample without additional thermal treatment. For comparison, we also include data for ZnO microparticles annealed in hydrogen ($H_2$) that were reported previously [5]. The preparation of that sample and the measurement conditions are described in [5].

Continuous-wave EPR measurements were performed at X-band (~9.4 GHz) using a Bruker ELEXSYS E580 spectrometer with an ER 4112HV variable-temperature helium-flow cryostat and an ER 4122 SHQE SuperX high-Q cavity. We performed EPR measurements at 0.3768 mW (26 dB) microwave power, a modulation frequency of 100 kHz, a modulation amplitude of 0.1 mT, a conversion time of 120 ms, and a spectral resolution of 2048 points. The ZnO powder samples were placed in thin-walled, clear, fused-quartz EPR sample tubes (Wilmad) with an outer diameter of 3 mm and filled to a height of approximately 10 mm. Comparable sample volumes and packing conditions were maintained for all measurements. We used standard DPPH free radical ($g$ = 2.0036) as a reference. The first derivative of the EPR spectra was simulated in the EasySpin 6.0.12 software package [12]. We reproduced the simulated spectra using Lorentzian and Voigt lineshape components via the ‘pepper’ function to extract spectral parameters, such as resonance field, linewidth, and relative intensity. We determined the microwave resonator quality factor $Q$ from the resonance response at a fixed attenuation of 33 dB.

We measured temperature-dependent EPR spectra from 5 to 297 K to investigate the thermal evolution of paramagnetic centers in the initial and annealed ZnO microparticles. We analyzed the experimental spectra by spectral decomposition to distinguish contributions from individual paramagnetic centers and determine their characteristic $g$-factors. For axially symmetric centers, the principal components $g_\perp$ and $g_\parallel$ were determined, and the corresponding average $g$-factor was calculated as $g_{av} = (2g_\perp + g_\parallel)/3$.

To analyze the temperature evolution of the paramagnetic-center populations, we double-integrated the simulated EPR signal components. The resulting double-integrated intensity, $DI$, was corrected for the temperature dependence of the resonator quality factor, $Q(T)$, normalized to its room-temperature value, and multiplied by temperature to account for the Curie-law contribution: $DI_{corr}(T) = DI(T)\ T\ Q(297\text{K})/Q(T)$.

### 3. Results and discussion

Fig. 1 compares the EPR spectra of the initial and annealed ZnO microparticles at 5 K. The initial ZnO shows a weak EPR signal with an overall spectral shape similar to the $N_2$-annealed sample. However, its low intensity prevents reliable decomposition into individual components. Annealing strongly enhances the EPR response, and differences among the $N_2$-, air-, and $H_2$-annealed spectra show that the paramagnetic defect population depends strongly on the annealing atmosphere. The EPR spectrum of $H_2$-annealed sample, previously investigated in [5], is included for comparison.

Fig. 2 shows the evolution of the EPR spectra of the $N_2$- and air-annealed ZnO microparticles over 5–297 K. Pronounced changes in both the overall signal amplitude and EPR spectral lineshape are observed with increasing temperature. In particular, the number and relative contributions of the spectral components change substantially, indicating different temperature dependences of the paramagnetic centers contributing to the EPR spectra. These changes are especially pronounced below approximately 50-60 K, where several low-temperature components are strongly suppressed.

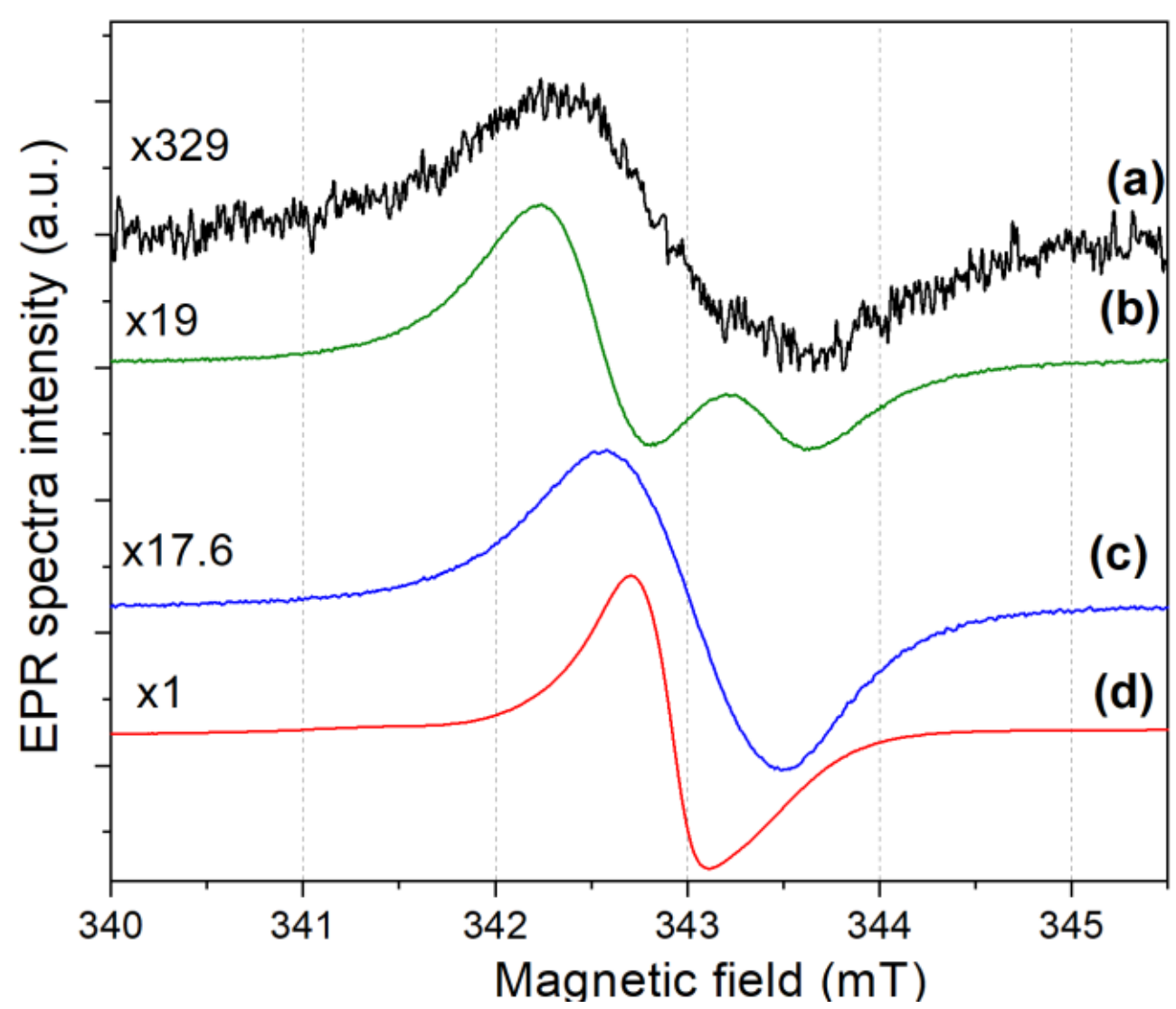

**Fig. 1.** X-band EPR spectra of the initial ZnO microparticles (a) and ZnO microparticles annealed in $N_2$ (b), air (c), and $H_2$ (d) recorded at 5 K. The $H_2$-annealed spectrum was taken from [5]. The scaling factors are indicated for each spectrum; therefore, the displayed amplitudes should not be compared quantitatively.

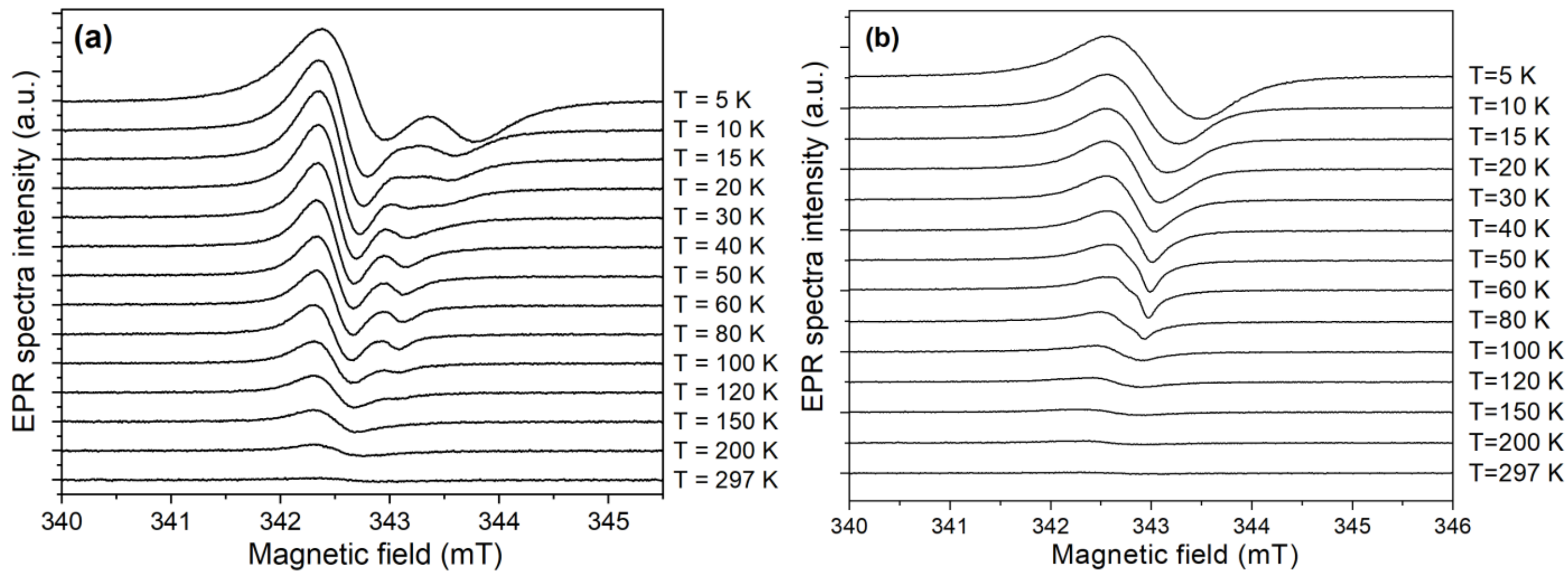

**Fig. 2.** Temperature dependence of EPR spectra from ZnO microparticles annealed in $N_2$ (a) and air (b).

To distinguish the individual EPR spectral components and follow their temperature behavior, the experimental EPR spectra were simulated at each temperature for both samples. Representative results at 5 K and 50 K are shown in Fig. 3. The resolved components were assigned by comparing their *g*-factor values and spectral characteristics with those previously identified in $H_2$-annealed ZnO microparticles [5], as summarized together with the obtained spin-Hamiltonian parameters in Table 1. It is seen that the $N_2$-annealed sample contains contributions from $Zn_i$, $V_O^+$, and shallow-donor centers, whereas the air-annealed sample contains $Zn_i$, $V_O^+$, and $H_i$-related centers. Thus, although $Zn_i$ and $V_O^+$-related centers are common to both air- and $N_2$-annealed samples, the additional low-temperature donor-related component depends on the annealing atmosphere: a shallow-donor component is observed after $N_2$ annealing, whereas an $H_i$-related component is observed after annealing in air.

Fig. 4 shows the temperature dependence of the double-integrated intensities of the individual spectral components obtained from the EPR spectral simulations. For the $N_2$-annealed sample, the shallow-donor contribution rapidly decreases with increasing temperature and becomes negligible above ~50 K. At the same time, the $Zn_i$ contribution initially increases, peaks around 40–50 K, and then decreases. In contrast, the $V_O^+$ contribution generally increases with temperature and remains observable up to room temperature. For the air-annealed sample, the $H_i$ contribution decreases rapidly with increasing temperature and becomes negligible at ~40-50 K. The $Zn_i$ contribution also shows a non-monotonic temperature dependence, whereas the $V_O^+$ contribution increases with temperature and tends to saturate at higher temperatures.

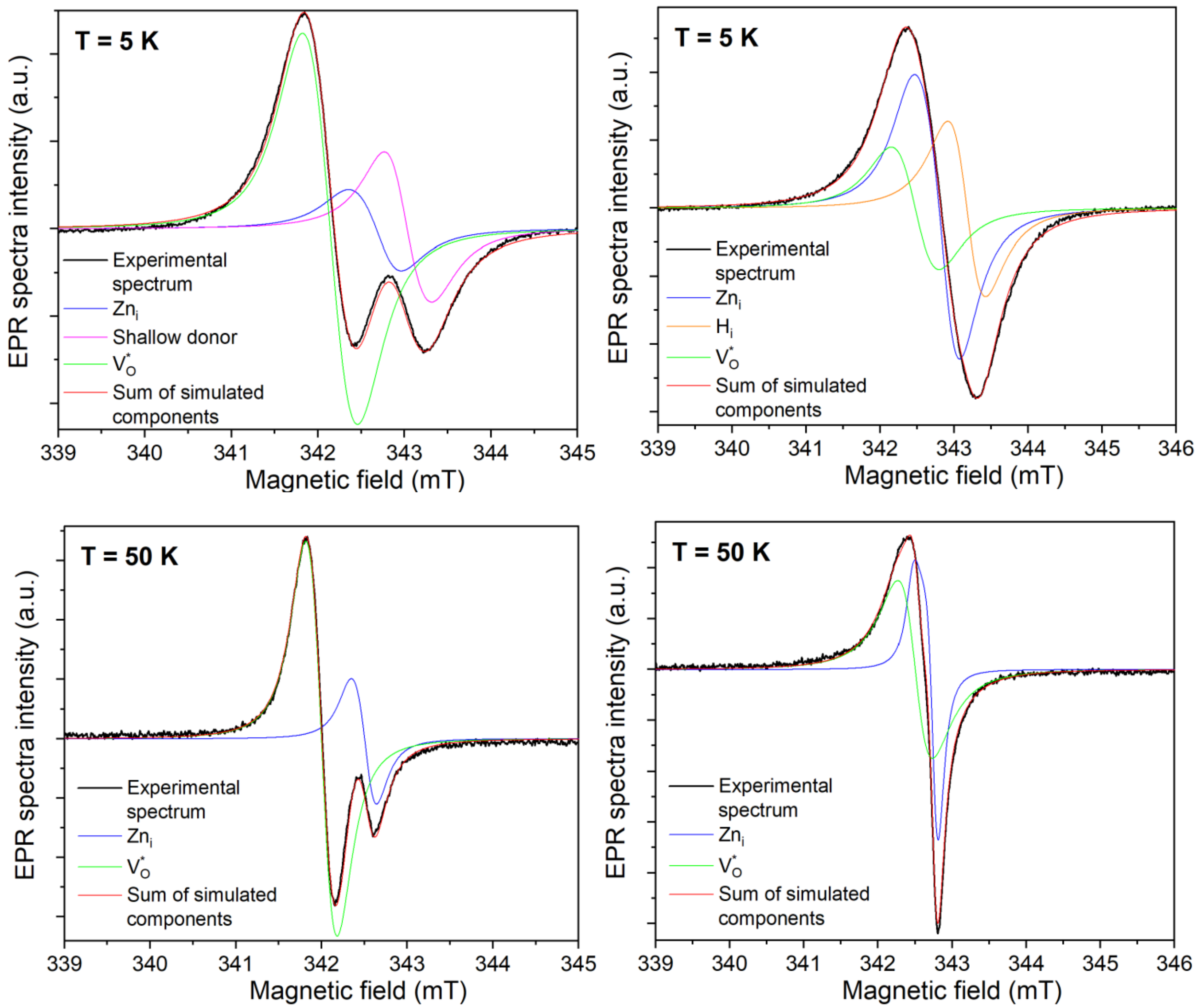

**Fig. 3.** Experimental EPR spectra (black curves) and their simulated components for ZnO microparticles annealed in $N_2$ (left) and air (right), recorded at 5 and 50 K. The individual components are assigned to $Zn_i$ (blue curves), $V_O^+$ (green curves), $H_i$ (orange curves), and shallow-donor (magenta curves) centers. The red curves represent the sums of the simulated components.

The different temperature dependences of the EPR spectra components of ZnO microparticles reflect different electronic properties of the observed defect centers. The monotonic quenching of the shallow-donor and $H_i$ EPR signals, which become negligible at ~50 K, is consistent with thermal depopulation of weakly localized donor states, in agreement with previous spectroscopic studies of H-related shallow donors in ZnO [9,14].

**Table 1.** Spin-Hamiltonian parameters of the paramagnetic centers observed in ZnO microparticles annealed in air and $N_2$ atmospheres, compared with those previously reported for $H_2$-annealed ZnO microparticles [5].

| Center | $g_\perp$ | $g_\parallel$ | $g_{av}$ | Annealing atmosphere | Reference |
|---|---|---|---|---|---|
| $V_O^+$ | 1.9623(5) | 1.9623(5) | 1.9623(5) | $H_2$ | [5] |
| | 1.9597(5) | 1.9597(5) | 1.9597(5) | Air | This work |
| | 1.9605(5) | 1.9605(5) | 1.9605(5) | $N_2$ | This work |
| $Zn_i$ | 1.9562(5) | 1.9581(5) | 1.9568(5) | $H_2$ | [5] |
| | 1.9570(5) | 1.9598(5) | 1.9580(5) | Air | This work |
| | 1.9567(5) | 1.9591(5) | 1.9575(5) | $N_2$ | This work |
| $H_i$ | 1.9556(5) | 1.9567(5) | 1.9559(5) | $H_2$ | [5] |
| | 1.9553(5) | 1.9566(5) | 1.9557(5) | Air | This work |
| Shallow donor | 1.9556(5) | 1.9538(5) | 1.9550(5) | $H_2$ | [5] |
| | 1.9560(5) | 1.9540(5) | 1.9553(5) | $N_2$ | This work |

A different behavior is observed for the $V_O^+$ center, whose EPR intensity does not show thermal quenching up to 297 K. Instead, its population increases with temperature. One possible explanation is that electrons released by thermal ionization of shallower donor centers are captured by initially EPR-silent $V_O^{2+}$ centers, increasing the population of paramagnetic $V_O^+$. This behavior is consistent with the deep-donor character of $V_O$ in ZnO and with our previous identification of $V_O^+$ as a compensating center involved in impurity-band conduction [5]. Photoluminescence studies also support the deep nature of $V_O$, associating the green emission band near 2.45 eV with a $V_O$ donor level about 530 meV below the conduction band and reporting a metastable $V_O^{2+}$ level at about 140 meV [13].

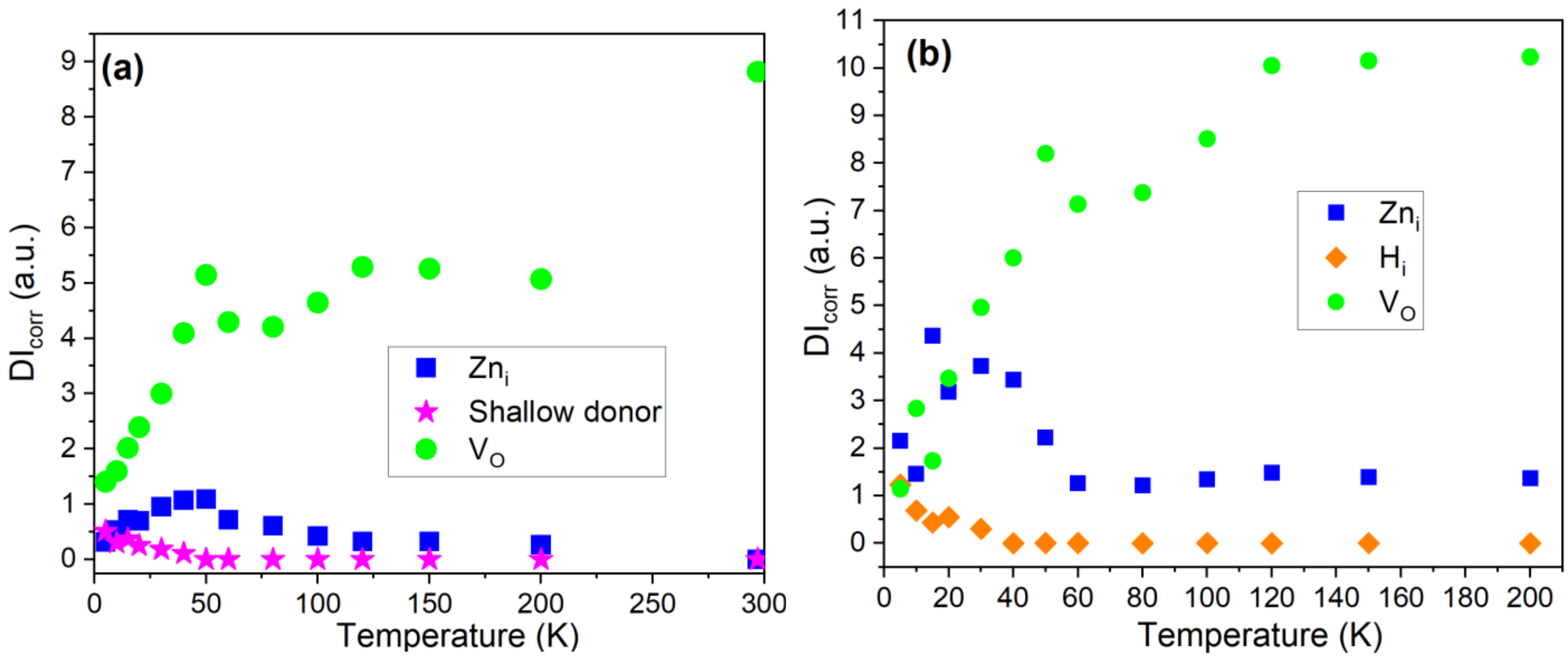

**Fig. 4.** Temperature dependence of the corrected double-integrated EPR intensities of the individual spectral components of $Zn_i$ (solid blue squares), $V_O^+$ (solid green circles), $H_i$ (solid orange rhombi), and shallow donor (solid magenta stars) centers obtained from the simulation of the EPR spectra of ZnO microparticles annealed in $N_2$ (a) and air (b) atmospheres.

Thus, the annealing atmosphere does not simply change the concentration of paramagnetic defects in ZnO microparticles but redistributes the population among donor states with different degrees of electron localization. In both $N_2$- and air-annealed samples, the shallow donor-related components are rapidly suppressed, while $V_O^+$ persists and increases in contribution at higher temperatures, consistent with a temperature-driven redistribution of electrons from weakly localized donor states toward deeper $V_O$-related states. At the same time, the different low-temperature components observed after $N_2$ and air annealing of ZnO microparticles show that the atmosphere determines which donor states participate in this redistribution: a shallow-donor center is observed after $N_2$ annealing, whereas an $H_i$-related center is characteristic of the air-annealed sample.

Fig. 5 shows the temperature dependence of the cavity $Q$-factor for the initial, $N_2$-annealed, and air-annealed ZnO microparticles. For comparison, the $Q(T)$ dependence previously obtained for $H_2$-annealed ZnO microparticles [5] is also included. The $Q$-factor value of the $N_2$- and air-annealed samples generally decreases with increasing temperature, similar to the behavior previously observed in $H_2$-annealed ZnO microparticles, although the decrease is considerably more pronounced for the $H_2$-annealed sample. In contrast, the initial ZnO powder shows an overall increase in the $Q$-factor with temperature. The contrasting $Q(T)$ behavior of the initial and annealed samples indicates that annealing significantly modifies the temperature-dependent microwave losses in ZnO.

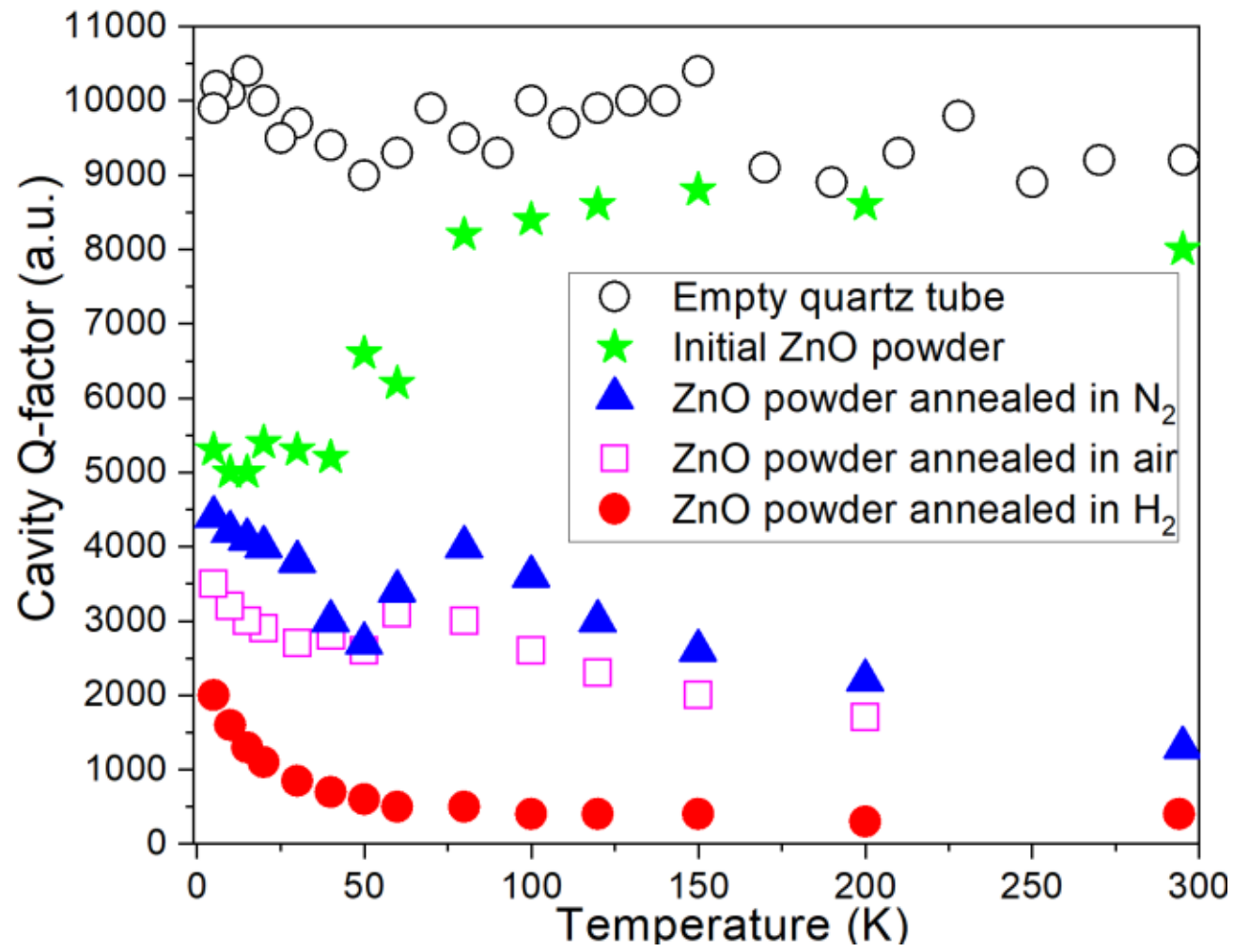

**Fig. 5.** Temperature dependence of the cavity *Q*-factor for the empty quartz tube (open black circles), initial ZnO microparticles (solid green stars), and ZnO microparticles annealed in $N_2$ (solid blue triangles) and air (open magenta squares). Data for $H_2$-annealed ZnO microparticles (solid red circles) from our previous study [5] are included for comparison.

Notably, the $N_2$- and air-annealed samples exhibit comparable *Q*-factor values despite their different paramagnetic defect compositions. This suggests that the microwave response cannot be associated with a single EPR-active defect center but likely reflects the combined influence of different donor-related defects and their temperature-dependent charge states.

The atmosphere-dependent differences in the paramagnetic defect structure and relative microwave conductivity are summarized in Table 2.

**Table 2.** Summary of the paramagnetic centers and relative microwave conductivity level of ZnO microparticles under different annealing conditions.

| Sample | Main paramagnetic centers | Relative microwave conductivity* |
|---|---|---|
| Initial ZnO | $Zn_i$ and $V_O^+$ likely present; individual components not reliably resolved | Low |
| $N_2$-annealed | $Zn_i$, $V_O^+$, shallow donor | Moderate |
| Air-annealed | $Zn_i$, $V_O^+$, $H_i$-related center | Moderate |
| $H_2$-annealed | $Zn_i$, $V_O^+$, $H_i$-related center, shallow donor | High |

**The relative microwave conductivity is qualitatively inferred from the cavity Q-factor and does not represent a direct conductivity measurement*

## 4. Conclusions

High-temperature annealing in $N_2$ and air produces distinctly different paramagnetic defects in ZnO microparticles. Temperature-dependent EPR spectroscopy combined with spectral simulations revealed contributions from $Zn_i$, $V_O^+$, and shallow-donor centers in the $N_2$-annealed sample, whereas $Zn_i$, $V_O^+$, and $H_i$-related centers were identified after annealing in air.

The resolved paramagnetic centers exhibit markedly different temperature dependences. The shallow-donor and $H_i$ signals are rapidly quenched and become negligible at ~50 K, consistent with thermal depopulation of weakly localized donor states. The $Zn_i$ contribution shows nonmonotonic behavior, increasing initially and then quenching at higher temperatures. In contrast, the $V_O^+$ signal remains EPR-active up to room temperature and increases with temperature. This behavior is consistent with the deep-donor character of $V_O$ in ZnO and may be explained by electron redistribution from thermally ionized shallower donor centers to initially EPR-silent $V_O^{2+}$ centers, forming paramagnetic $V_O^+$.

The cavity *Q*-factor measurements further show that annealing significantly modifies the temperature-dependent microwave losses in ZnO. The $N_2$- and air-annealed samples show comparable *Q*-factor values despite different paramagnetic defect compositions, suggesting the microwave response is not associated with a single EPR-active defect center.

Our results demonstrate that the annealing atmosphere controls the composition, charge-state population, and thermal behavior of paramagnetic defects and affects the associated electronic response of ZnO microparticles. These findings further clarify the relationship between donor-related defects, charge redistribution, and the electronic response of ZnO. From a practical perspective, controlling the defect structure of ZnO microparticles through the annealing atmosphere may be useful when these particles are used as precursors for ZnO films and coatings, allowing their electronic properties to be adjusted at the precursor-processing stage.

**CRediT authorship contribution statement**

**D. Savchenko:** Data curation, Formal analysis, Investigation, Methodology, Visualization, Writing - original draft, Writing - review & editing. **M. Holiatkina:** Data curation, Formal analysis, Software, Visualization, Writing - original draft. **A. Prokhorov:** Investigation, Writing - review & editing. **A. Vasin:** Methodology, Validation, Writing - review & editing. **E. Kalabukhova:** Conceptualization, Supervision, Validation, Writing - original draft, Writing - review & editing.

**Declaration of Generative AI and AI-assisted technologies in the writing process**

During the preparation of this work, the authors used Grammarly in order to review spelling and grammar. The tool was not used for generating scientific content, analysis, or results. After using this tool/service, the authors reviewed and edited the content as needed and take full responsibility for the content of the publication.

**Declaration of Competing Interest**

The authors declare that they have no known competing financial interests or personal relationships that could have influenced the work reported in this paper.